\documentclass[twoside]{article}
\usepackage{fleqn,espcrc2,axodraw}

\usepackage{graphicx}

\newcommand{\AmS}{{\protect\the\textfont2
  A\kern-.1667em\lower.5ex\hbox{M}\kern-.125emS}}
\newcommand{\half}{\textstyle{1\over2}}

\begin{document}

\title{
\vspace*{-13mm} \hfill {\small CTS-IISc-19/01}\\
\vspace*{-3mm}  \hfill {\small hep-lat/0111046}\\
Mesons in transverse lattice QCD at strong coupling and large-N%
\thanks{Presented at ``Lattice 2001'', 19-24 August 2001, Berlin, Germany.
        To appear in the proceedings.}
      }

\author{Apoorva D. Patel\address{CTS and SERC, Indian Institute of Science,
        Bangalore-560012, India}
        (e-mail: adpatel@cts.iisc.ernet.in)
       }
       
\begin{abstract}
Mesons in large-N QCD are analysed in light-front coordinates with a
transverse lattice at strong coupling. In this limit, their properties
can be expressed as simple renormalisations of the 't Hooft model results.
The integral eigenvalue equation for the mesons is derived. Spectrum,
lightcone wavefunctions and form factors of various mesons can be
numerically calculated from this equation.
\end{abstract}

\maketitle

I have proposed that large-N QCD, with two continuous dimensions on the
lightcone and two transverse dimensions on lattice at strong coupling,
can be a useful phenomenological model of certain hadronic properties
\cite{PATEL}.
This combination of large-N and strong coupling limits is chosen to
obtain analytically tractable results. The large-N limit allows the use
of exact results for two dimensional QCD \cite{THOOFT},
while the strong coupling limit permits systematic incorporation of the
effects of additional transverse directions.

Let $\mu,\nu$ denote the lightcone directions, and $i,j$ the lattice
directions. The action for the theory is ($g$ is held fixed as
$N\rightarrow\infty$):
\begin{eqnarray}
S &=& {a_\perp^2 N \over g^2}\sum_{x_\perp}\int d^2x~ \Big[
   -  {\half} \sum_{\mu,\nu}
      {\rm Tr} \left( F_{\mu\nu}(x) F^{\mu\nu}(x) \right) \nonumber\\
  &+& \overline{\psi}(x) (i\sum_\mu \gamma^\mu \partial_\mu
                         - \sum_\mu \gamma^\mu A_\mu - m) \psi(x) \nonumber\\
  &+& i\kappa \sum_i \big\{
      \overline{\psi}(x) (r+\gamma^i) U_i(x) \psi(x+\hat{i}a_\perp) \nonumber\\
  &+& \overline{\psi}(x) (r-\gamma^i) U_i^\dag(x-\hat{i}a_\perp)
                                          \psi(x-\hat{i}a_\perp) \big\} \Big]
\end{eqnarray}
This action already incorporates $g_\perp\rightarrow\infty$, for the gauge
coupling corresponding to the transverse directions. The transverse lattice
spacing, $a_\perp$, remains finite in this limit. The terms,
${\rm Tr}(U_i(x) U_j(x+\hat{i}a_\perp) U^{i\dag}(x+\hat{j}a_\perp) U^{j\dag}(x))$
and ${\rm Tr}(\partial_\mu U_i(x) \partial^\mu U^{i\dag}(x+\hat{i}a_\perp))$,
are allowed by symmetry considerations, but they have been dropped because
they are suppressed by inverse powers of $g_\perp$. The functional integral
over $U_\perp(x)$ reduces to the constraint that in any correlation function
products of $U_\perp(x)$ must contract to a colour singlet at each space-time
point.

The specification of lattice fermions is not unique, and here I have chosen
the nearest neighbour discretisation. $r=0,1$ correspond to naive and Wilson
fermions respectively. At weak coupling $\kappa=1/(2a_\perp)$, but at strong
coupling its value is different due to non-perturbative renormalisations.
The parameters of the action, that have to be fitted to physical results,
are therefore $g^2/a_\perp^2$, $m$ and $\kappa$.

In this theory, the choice of axial gauge $A^+=0$ reduces the lightcone
gauge field dynamics to a linear confining potential, just as in the case of
the 't Hooft model. Unlike the 't Hooft model, however, the $\gamma-$matrices
cannot be eliminated from fermion propagators and external sources. They
become the spin-parity labels of hadron states.

The interplay between large-N and chiral limits is subtle in the 't Hooft
model \cite{ZHITNITSKY}.
The chiral symmetry is realised in the Berezinski\u\i-Kosterlitz-Thouless
mode in this two-dimensional theory:
$\langle\overline\psi\psi(x) \overline\psi\psi(0)\rangle \sim x^{-1/N}$.
The large-N limit must be taken before taking the chiral limit to obtain
results relevant to QCD. Contributions from graphs involving non-planar
pions are suppressed by inverse powers of N, but are enhanced by inverse
powers of $m_\pi$. If the chiral limit is taken before the large-N limit,
one obtains the wrong phase of the theory with massless baryons. If the
large-N limit is taken first, then the baryons become infinitely massive
solitons and it is not easy to obtain results for them. Properties of
mesons can be extracted in a straightforward manner, however, following
the analysis of 't Hooft.

The large-N and transverse strong coupling limits separate the hadron Green
functions into two types of non-overlapping factors---propagators along the
lightcone and hops along the transverse lattice \cite{PATEL}.
The two factors can be evaluated separately and then combined to yield the
final results.

The first step is to obtain the renormalised quark propagator. For the
quark propagator along the lightcone, the self-energy term consists of
two contributions depicted in Fig.1. (a) is the contribution evaluated by
't Hooft,
\begin{equation}
\Sigma_{2d}(p) = -{g^2 \gamma^+ \over 2\pi a_\perp^2 p_-} ~.
\end{equation}
(b) is the new contribution due to the transverse lattice---it is a scalar
tadpole term, $T$. Both types of insertions are easily summed up as geometric
series. The sum of tadpole insertions renormalises the quark mass, and the
sum of gluon corrections shifts the pole position. The quark propagator along
the lightcone thus becomes
\begin{eqnarray}
S(p) &=& {i \over p\!\!\!/ - \Sigma_{2d}(p) - m - T} \\
     &=& {i ( p\!\!\!/ - \Sigma_{2d}(p) + m + T )
         \over 2p_+p_- - (m+T)^2 + (g^2/\pi a_\perp^2)} ~.
\end{eqnarray}
The lightcone $\gamma-$matrix algebra implies that the part of $S(p)$
proportional to $\gamma^-$ is the most important for calculations of
hadron Green functions.

\begin{picture}(200,90)(0,0)
\Text(40,5)[c]{$(a)$}
\ArrowLine(0,20)(20,20) \Line(20,20)(60,20) \ArrowLine(60,20)(80,20)
\Vertex(40,20){4} \PhotonArc(40,20)(24,0,180){2}{10}
\Text(150,5)[c]{$(b)$}
\ArrowLine(110,20)(150,20) \ArrowLine(154,20)(194,20)
\Line(150,20)(150,40) \Line(154,20)(154,40)
\CArc(152,56)(16,278,262) \Vertex(152,72){4}
\end{picture}

\smallskip
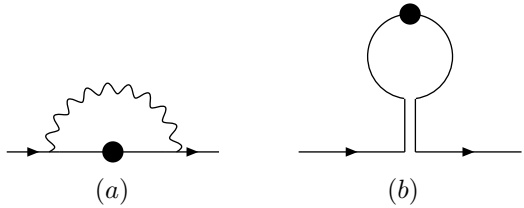
\noindent Figure 1. Self energy contributions to the quark propagator:
(a) gluonic correction from the continuous dimensions,
(b) tadpole correction due to hops in the lattice directions.
Fermion line with a filled dot denotes the full quark propagator.
\medskip

Quark hops along the transverse lattice contribute the factors $i\kappa
(r\pm\gamma^i) \exp(\pm i p_i a_\perp)$ to the propagator. These hops are
also renormalised by tadpole insertions, which can be taken into account
by replacing $\kappa$ with $\widetilde\kappa \equiv \kappa/(1-T_\perp)$.
Both $T$ and $T_\perp$ are calculable functions of the parameters in the
action. But they always appear in particular combinations, and so one can
parametrise the results in terms of $g^2/a_\perp^2$, $\widetilde{m} \equiv
m+T$ and $\widetilde\kappa$.
(It can be noted that the tadpole corrections vanish for Wilson fermions
due to their projection operator spin propagator structure.)

The meson propagator is obtained by including interactions between the
renormalised quark and antiquark propagators. There are two types of
interactions, as shown in Fig.2. (a) is the gluon exchange in the
continuous dimensions, and 't Hooft showed that it produces a linear
confining potential between the quark and the antiquark.
\begin{equation}
V_{2d} = - i {g^2 \over a_\perp^2} {P \over (k_-)^2}
         \cdot\left[ \gamma^+ \otimes (\gamma^+)^T \right]
\end{equation}
Here $k$ is the gluon momentum and the direct product sign separates the
spin factors acting on quark and antiquark propagators. (b) is the new
interaction which collapses the quark and the antiquark into a colour
singlet meson and then hops the meson along the lattice directions.
\begin{eqnarray}
V_\perp = && - \widetilde\kappa^2 \sum_{i,\pm} \delta(x_\mu-y_\mu)
          ~ \exp(\pm i p_i a_\perp) \nonumber\\
          &&\cdot\left[ (r\pm\gamma^i) \otimes (r\mp\gamma^i)^T \right] ~,
\end{eqnarray}
where $x$ and $y$ are quark and antiquark positions, and $p$ is the meson
momentum. The position space $\delta-$function constraint translates into
a momentum space interaction that is independent of the relative momentum
$k_\mu$.

\smallskip
\begin{picture}(200,75)(0,0)
\Text(40,5)[c]{$(a)$}
\ArrowLine(0,20)(20,20) \Line(20,20)(60,20) \ArrowLine(60,20)(80,20)
\Vertex(20,20){4} \Vertex(60,20){4} 
\ArrowLine(80,60)(60,60) \Line(60,60)(20,60) \ArrowLine(20,60)(0,60)
\Vertex(20,60){4} \Vertex(60,60){4} 
\Photon(40,20)(40,60){2}{6}
\Text(150,5)[c]{$(b)$}
\ArrowLine(110,20)(130,32) \Line(130,32)(140,38) \Line(140,38)(160,38)
\Line(160,38)(170,32) \ArrowLine(170,32)(190,20)
\Vertex(130,32){4} \Vertex(170,32){4} 
\ArrowLine(190,60)(170,48) \Line(170,48)(160,42) \Line(160,42)(140,42)
\Line(140,42)(130,48) \ArrowLine(130,48)(110,60)
\Vertex(130,48){4} \Vertex(170,48){4} 
\end{picture}

\medskip
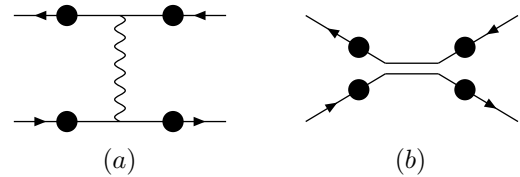
\noindent Figure 2. The interactions between quark and anti-quark in a
meson: (a) gluon exchange in the continuous dimensions,
(b) colour singlet hops in the lattice directions.
Fermion line with a filled dot denotes the full quark propagator.
\medskip

The two types of interactions do not mix with each other, and just produce
a series of ladder diagrams. The properties of the meson poles of the Green
functions are easily extracted from a Bethe-Salpeter equation, which sums up
the ladder diagrams. For a meson state with spin-parity structure $\Gamma$,
let
\begin{equation}
\psi_\Gamma(p,q) = \langle \overline{\psi}(p-q) \Gamma \psi(q)
                 | {\rm Meson}_\Gamma (p) \rangle ~,
\end{equation}
\begin{equation}
\phi_\Gamma(p,q_-) = \int d^2q_\perp dq_+ ~ \psi_\Gamma(p,q) ~.
\end{equation}
Only $\phi_\Gamma(p,q_-)$ appear in the Bethe-Salpeter equation, because
the interactions in Eqs.(5,6) do not depend on $q_\perp,q_+$. Indeed
$\phi_\Gamma(p,q_-)$ are the lightcone wavefunctions of the quarks in the
mesons. They satisfy the integral equations
\begin{eqnarray}
\Gamma \phi_\Gamma(p,q_-) &=& \int {dq_+\over2\pi} ~ S(p-q) \Gamma S(-q)
  \nonumber\\
  &\cdot& \int {dk_-\over2\pi} \left[ V_{2d} + V_\perp \right]
  \phi_\Gamma(p,k_-) ~.
\end{eqnarray}
It is implicitly understood here that the spin factors are ordered along the
fermion line. In general, the equations for different spin-parity structures
$\Gamma$ are coupled. The eigenstates can be found by diagonalising the
$\Gamma-$dependent integral equations, as in the case of conventional strong
coupling expansions \cite{FRIEDBERG}.
Integral equations for mesons with unequal quark masses are obtained by
inserting appropriate quark masses in the propagators $S$ and replacing
$\widetilde\kappa^2$ by $\widetilde\kappa_1 \widetilde\kappa_2$.

The transverse lattice directions explicitly break the rotational symmetry.
Still parity is a good quantum number, and helicity can be defined modulo
4 on the hypercubic lattice. They restrict the mixing amongst various
$\Gamma-$structures. While perfect rotational symmetry cannot be realised,
the free parameters of the theory, $g^2/a_\perp^2$ and $\widetilde\kappa$,
can be non-perturbatively fixed, by demanding that (i) various helicity
states belonging to the same angular momentum multiplet be as degenerate
as possible, and (ii) the dispersion relation for mesons be as close to
the rotationally symmetric situation as possible.

Even without explicitly solving the integral equations, it can be inferred
that the meson spectrum consists of a tower of states in each quantum
number channel, and the towers of states with different spin form a set of
parallel trajectories \cite{PATEL}. 
The separations between trajectories depend on the type of transverse
lattice discretisation of fermions. Such a behaviour is expected from a
successful combination of the 't Hooft model results and strong coupling
features.

Explicitly, the integral over $q_+$ in Eq.(9) was evaluated by 't Hooft,
and the integral of the new interaction $V_\perp$ over $k_-$ is just a
constant. The numerical value of the constant depends on $\Gamma$ through
the commutation properties of various spin factors, and so the meson poles
for different $\Gamma-$structures are shifted with respect to each other.
(From results of conventional strong coupling expansions one can guess
that the shifts are essentially independent of the quark masses.)

The actual solutions for the meson spectrum and $\phi_\Gamma$ have to be
obtained numerically. The usefulness of the extreme transverse strong
coupling and large-N limit of QCD can only be judged by how well the
results fit the experimental data. I expect the formalism to provide
a good phenomenological description of deep inelastic scattering, because
there the transverse dimensions contribute only at subleading order and
so it should not matter much if they are highly distorted by a coarse
lattice. The methodology to extract structure functions and form factors
from moments of $\phi_\Gamma$ already exists in the case of 't Hooft
model \cite{CALLAN,EINHORN},
and it is straightforward to extend that to the transverse lattice geometry.
Detailed numerical analysis of all this is in progress.

\end{document}